# Search for macroscopic CP violating forces using a neutron EDM spectrometer

A.P. Serebrov [1,*], O. Zimmer [2], P. Geltenbort [2], A.K. Fomin [1],
S.N. Ivanov [1], E.A. Kolomensky [1], I.A. Krasnoshekova [1], M.S. Lasakov [1],
V.M. Lobashev [1], A.N. Pirozhkov [1], V.E. Varlamov [1], A.V. Vasiliev [1],
O.M. Zherebtsov [1], E.B. Aleksandrov [3], S.P. Dmitriev [3], N.A. Dovator [3],

[1] *Petersburg Nuclear Physics Institute, RAS, 188300, Gatchina, Leningrad District, Russia*

[2] *Institut Laue Langevin, BP 156, 38042 Grenoble cedex 9, France*

[3] *Ioffe Physico-Technical Institute, RAS, 194021, St. Petersburg, Russia*

---

[*] *Corresponding author:* A.P. Serebrov

A.P. Serebrov

Petersburg Nuclear Physics Institute

Gatchina, Leningrad district

188300 Russia

Telephone: +7 81371 46001

Fax: +7 81371 30072

E-mail: serebrov@pnpi.spb.ru

**Abstract**

The search for CP violating forces between nucleons in the so-called axion window of force ranges $\lambda$ between $2\times10^{-5}$ m and $0.02$ m is interesting because only little experimental information is available there. Axion-like particles would induce a pseudo-magnetic field for neutrons close to bulk matter. A laboratory search investigates neutron spin precession close to a heavy mirror using ultracold neutrons in a magnetic resonance spectrometer. From the absence of a shift of the magnetic resonance we established new constraints on the coupling strength of axion-like particles in terms of the product $g_s g_p$ of scalar and pseudo-scalar dimensionless constants, as a function of the force range $\lambda$, $g_s g_p \lambda^2 \leq 2\times10^{-21}$ [cm$^2$] (C.L.95%) for $10^{-4}$ cm $< \lambda <$ 1 cm. For 0.1 cm $< \lambda <$ 1 cm previous limits are improved by 4 to 5 orders of magnitude.

1. **Introduction**

Axions are hypothetic particles which were introduced to understand the absence of CP-violation in strong interactions between quarks [1,2]. Axions and axion-like particles [3] are considered as candidates for dark matter in the Universe. If existent they would mediate CP-violating spin-dependent forces between nucleons over macroscopic distances. The potential between a neutron (acting as probe) and a nucleon due to single-boson exchange between a scalar and a pseudo-scalar vertex with corresponding coupling constants $g_s$ and $g_p$ has the following form [4]:

$$V(\boldsymbol{r}) = \frac{\hbar^2}{8\pi m_n} g_s g_p \boldsymbol{n} \cdot \boldsymbol{\sigma}_n \left(\frac{1}{\lambda r} + \frac{1}{r^2}\right) e^{-r/\lambda} \qquad (1)$$

$\boldsymbol{n} = \boldsymbol{r}/r$ is a unit vector between neutron and nucleon, $\lambda = \hbar/(m_A c)$ is the range of forces (Compton wavelength of axion-like particle with mass $m_A$), $m_n$ is the neutron mass and $\boldsymbol{\sigma}_n$ is the vector of Pauli matrices belonging to the spin of the neutron. The correlation $\boldsymbol{n} \cdot \boldsymbol{\sigma}_n$ violates both P and T invariance. For a neutron at distance $l$ from a massive flat wall with thickness $d$ and nucleon number density $N$, the interaction (1) gives rise to a spin-dependent potential [5]:

$$V(l) = \pm \frac{\hbar^2 \lambda}{4 m_n} N g_s g_p e^{-l/\lambda} \left(1 - e^{-d/\lambda}\right) \qquad (2)$$

The sign is determined by the neutron spin projection onto the surface normal.

A spin-dependent potential gives rise to a spin precession with angular frequency $\omega = 2|V|/\hbar$. Ramsey's magnetic resonance technique is very sensitive to detect tiny precession angles and may be applied for detecting the pseudo-magnetic effect induced by axion-like particles on trapped neutrons [5]. In this paper we present results of first experiments using this method, taking advantage of an existing apparatus to search for the neutron electric dipole moment (EDM) [6], which is presently installed at a polarized beam of ultracold neutrons (UCN) [7] at the facility PF2 at the Institut Laue-Langevin in Grenoble, France.

2. Experiment

Two traps for UCN storage, with cylindrical shape and closed off with flat walls (see Fig. 1) are located in a magnetic guide field $\boldsymbol{H}_z$ with strength 2 µT, pointing in vertical z-direction, and shielded from external influences by four layers of µ-metal. The hypothetic spin-dependent force fields with a range short compared to the dimensions of the traps would be present only in the vicinity of the walls and lead to a local neutron spin precession in addition to the ordinary Larmor precession in the magnetic field. In presence of the spin-dependent potential (2) the horizontal walls will cause pseudo-magnetic fields parallel or anti-parallel to the magnetic guide field, which will result in a pseudo-magnetic precession angle for neutrons trapped in a superposition of spin states parallel and anti-parallel to the magnetic field. In order to obtain a large precession angle during neutron storage, partial cancellation effects due to pseudo-magnetic fields parallel and anti-parallel to the magnetic guide field have to be kept small. Therefore, in a trap with flat walls perpendicular to the magnetic field the wall plates have to be made from materials with a large difference in density. The central wall of the double storage trap shown in Fig. 1 is made from a thick Cu plate with density 8.9 g/cm$^3$, whereas the outer plates are made from aluminum with density 2.7 g/cm$^3$, coated with a 300 nm thin layer of beryllium. The total angle of pseudo-magnetic precession in each trap is found using the relation [5]

$$\varphi = \frac{2T}{\hbar D}\int_0^D V(z)\,\mathrm{d}z \quad (3)$$

where $D$ is the distance between the parallel plates of a trap and $T$ is the holding time of UCN (15 s in our experiment). Using eq. (2) we obtain

$$\varphi = \pm\frac{\hbar T}{2m_n}g_s g_p \frac{\lambda^2}{D}\left(1 - \mathrm{e}^{-D/\lambda}\right)\left[N_{\mathrm{Cu}}\left(1 - \mathrm{e}^{-d_{\mathrm{Cu}}/\lambda}\right) - N_{\mathrm{Al}}\left(1 - \mathrm{e}^{-d_{\mathrm{Al}}/\lambda}\right)\right] \quad (4)$$

where $N_i$ and $d_i$ are nucleon number densities and thicknesses of the corresponding trap plate. The sign in eq. (4) accounts for the opposite directions of the net pseudo-magnetic fields in the top and the bottom trap, leading to a relative phase shift

$$\varphi_{\text{rel}} = \varphi_{\text{top}} - \varphi_{\text{bottom}} = 2|\varphi| \qquad (5)$$

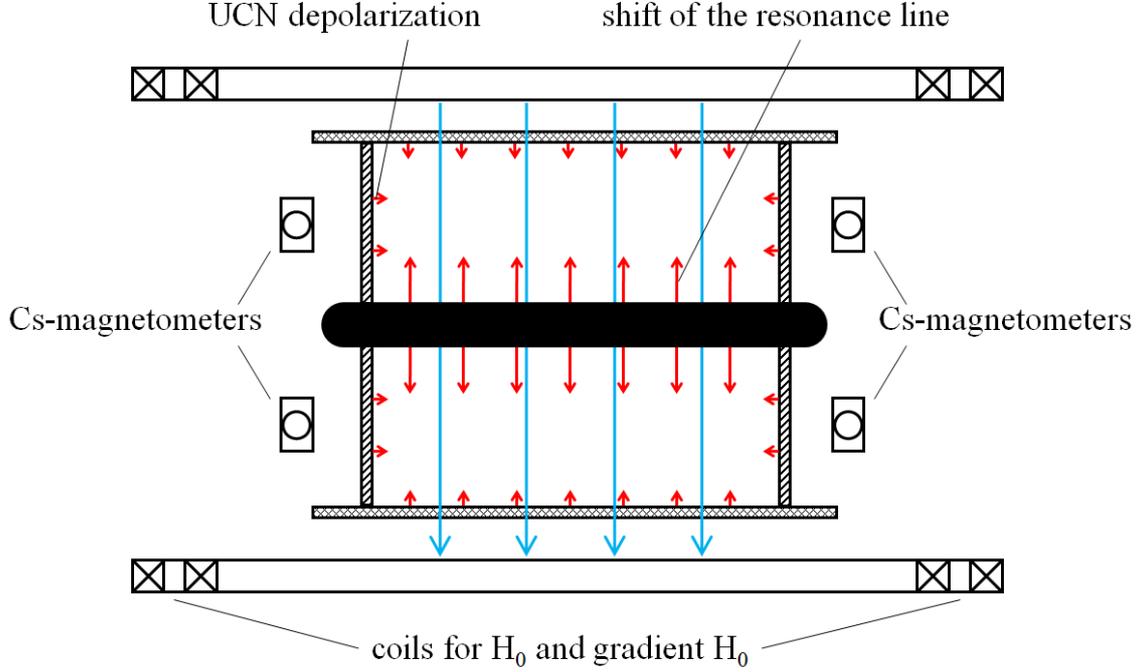

Figure 1: Experimental setup with double storage chamber for UCN, Cs-magnetometers and coils for field setting and correction.

However, such a shift may also occur for other reasons such as magnetic field gradients. The technically simplest method to distinguish the sought effect from a gradient-induced effect employs a reversal of the magnetic field without change of field gradients (see discussion further below) which will invert the sign of $\varphi_{\text{rel}}$ without inverting the sign of a gradient-induced shift. The difference of phase shifts for the two field directions will thus be given by

$$\varphi_+ - \varphi_- = 2\varphi_{\text{rel}} = 4|\varphi| \qquad (6)$$

Equation (4) with (6) tell us that we can express the sought product of coupling constants in terms of the measured angles and experimental parameters,

$$g_s g_p = \frac{\varphi_+ - \varphi_-}{\frac{2\hbar T}{m_n}\frac{\lambda^2}{D}(1 - e^{-D/\lambda})[N_{\text{Cu}}(1 - e^{-d_{\text{Cu}}/\lambda}) - N_{\text{Al}}(1 - e^{-d_{\text{Al}}/\lambda})]} \qquad (7)$$

The technical realization of the field reversal requires control of magnetic field strength and in particular of field gradients. To deal with the problem of hysteresis effects of the magnetic shielding of the spectrometer we employ eight Cs-magnetometers placed around the UCN traps (out of range of the sought forces induced by axion-like particles). Four magnetometers are situated in the plane of Fig. 1 and four further magnetometers in the orthogonal plane through the symmetry axis of the setup. The four upper (lower) Cs-magnetometers provide information about the magnetic field in the upper (lower) trap. In a first step of our measurements we match the resonances in both traps by tuning the current in a gradient coil system surrounding the traps (see Fig. 1). Figure 2a shows resonance curves matched this way for magnetic field pointing upwards with strength $H_+$. After field reversal ($H_-$) the currents in the main coils for magnetic field production and for field gradient correction were tuned to obtain (1) the same average value of magnetic field measured by the eight Cs-cells and (2) the same difference between average values for the four upper and the four lower magnetometers. For an axially symmetric magnetic field this procedure would be sufficient. Unfortunately, the magnetic shielding produces more complicated field gradients, which moreover are not reproducible. Figure 2b shows resonance curves after field reversal ($H_-$) and Fig. 2c after a second field reversal back to the original situation ($H_+$). The observed non-reproducibility severely limits the sensitivity of the field-reversal technique. To estimate the corresponding uncertainty of measurements field reversals were repeated many times.

Table 1 shows results of measured resonance shifts $\varphi_+^{Cu}$ and $\varphi_-^{Cu}$ between the traps for the two different directions of magnetic field ($H_+$ and $H_-$). The average value of the difference is $\langle\varphi_+^{Cu}\rangle - \langle\varphi_-^{Cu}\rangle = (0.70 \pm 3.83)$ rad. The uncertainty of the average values was calculated as mean squared deviation of results of individual measurements. Despite demagnetization of the magnetic screens before each new measurement after field reversal the results in Tab. 1 show a large dispersion due to residual hysteresis of the screens.

A similar series of experiments was performed with the central copper plate replaced by an aluminum plate, i.e. the same material as the outer trap walls. In this situation pseudo-magnetic fields from the central and outer trap walls compensate each other (a small correction due to difference of thickness of electrodes is taken into account in numerical calculations for final result). Table 2 shows the results of these measurements. $\langle\varphi_+^{Al}\rangle - \langle\varphi_-^{Al}\rangle = (-3.66 \pm 3.45)$ rad. For the pseudo-magnetic effect we obtain the final result:

$$\varphi_+ - \varphi_- = \langle\varphi_+^{Cu}\rangle - \langle\varphi_-^{Cu}\rangle - \left(\langle\varphi_+^{Al}\rangle - \langle\varphi_-^{Al}\rangle\right) = (4.36 \pm 5.14) \text{ rad} \qquad (8)$$

Interpreting this result as an upper limit on an effect induced by a pseudo-magnetic field we obtain $\varphi_+ - \varphi_- \leq 14.6$ rad (95% C.L.). The corresponding upper limit of the product $g_s g_p$ as function of $\lambda$ is shown in Fig. 3 by the line 1.

Table 1: Results of measurements with the central trap wall made from copper.

| N | $\varphi_+^{Cu}$ (rad) | $\varphi_+^{Cu}$ (rad) |
|---|---|---|
| 1 | 0.61 | 4.05 |
| 2 | 6.28 | 4.73 |
| 3 | 5.67 | 0.01 |
| 4 | 4.87 | 0.61 |
| 5 | 4.73 | 0.87 |
| 6 | 4.87 | 0.40 |
| 7 | 4.57 | 8.31 |
| 8 | 0.51 | 0.71 |
| 9 | -1.55 | 0.02 |
| 10 | 1.85 | |
| 11 | 3.14 | |
| | 3.23±2.53 | 2.53±2.87 |
| $\varphi_+^{Cu} - \varphi_-^{Cu} = (0.70 \pm 3.83)$ rad | | |

Table 2: Results of measurements with the central trap wall made from aluminum.

| N | $\varphi_+^{Al}$ (rad) | $\varphi_+^{Al}$ (rad) |
|---|---|---|
| 1 | 0.06 | 4.05 |
| 2 | -1.29 | 5.18 |
| 3 | -0.48 | 5.23 |
| 4 | 0.51 | -0.30 |
| 5 | 0.66 | 3.14 |
| 6 | 4.03 | 3.14 |
| 7 | -5.35 | |
| | 0.26±2.79 | 3.40±2.04 |
| $\varphi_+^{Al} - \varphi_-^{Al} = (-3.66 \pm 3.45)$ rad | | |

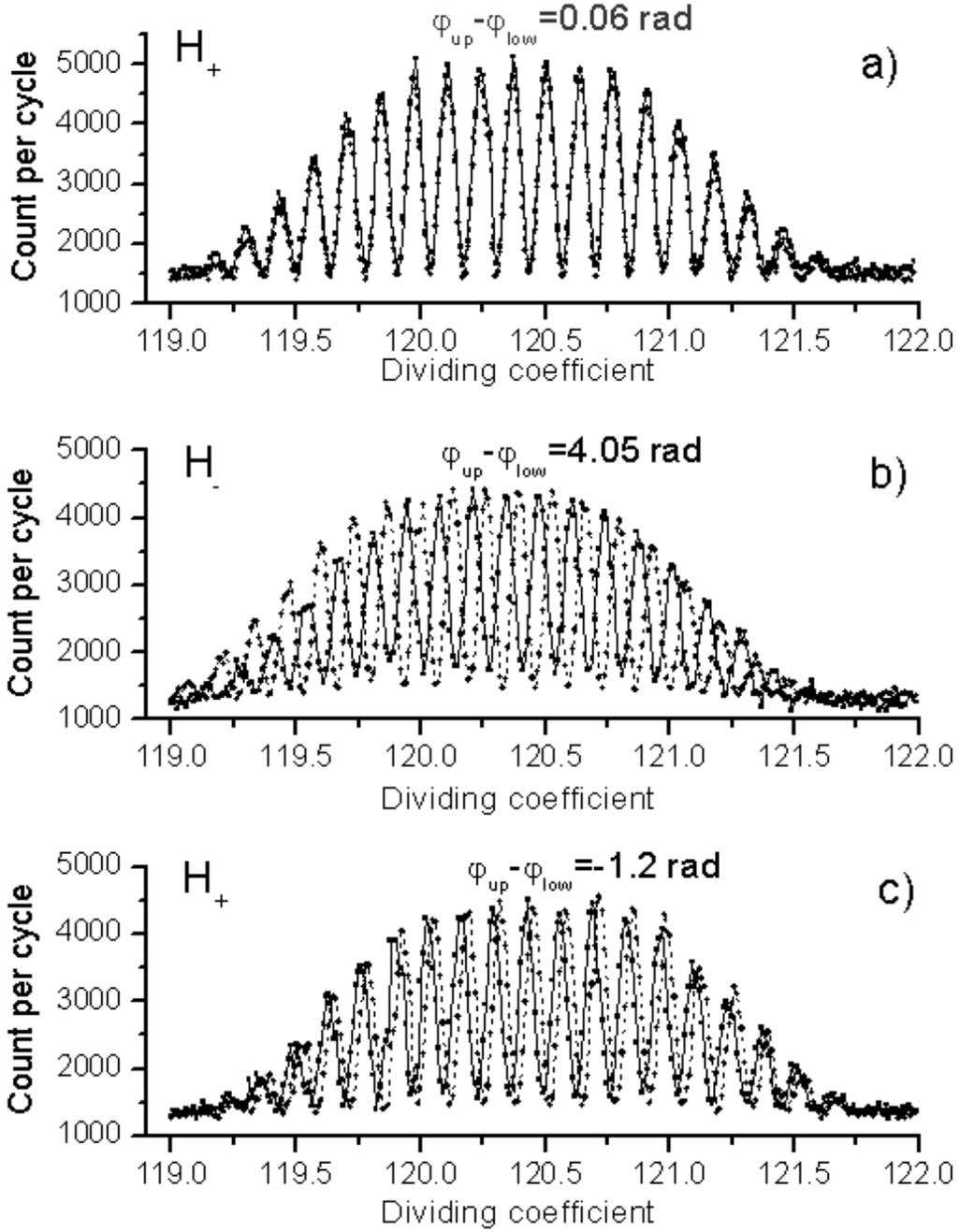

Figure 2: Resonance curves obtained in the double chamber EDM spectrometer using the chamber arrangement shown in Fig. 1. a) with magnetic field $H_+$ directed upwards, b) with magnetic field $H_-$ directed downwards, c) after a second field reversal to the original field direction $H_+$. The horizontal axis shows the coefficient by which the frequency of Cs-magnetometers has to be divided to obtain the frequency of neutron precession. The vertical axis shows the counts of neutron detector in the system of analysis of the UCN polarization after storage in the traps.

### 3. Discussion and conclusion

Our experiment has led to new constraints for the coupling constant product $g_s g_p$, as a function of the range $\lambda$ of the macroscopic force. The results are shown in Fig. 3 for studies of the shift of magnetic resonance (line 1). The obtained new constraints for $g_s g_p$ significantly improve a previous result deduced from the measured intensity of neutrons passing through a narrow channel formed by a horizontal mirror and an absorber ("gravitational levels" in Fig. 3) [8]. In the range 0.1 cm $\leq \lambda \leq$ 1 cm the improvement is 4 to 5 orders of magnitude.

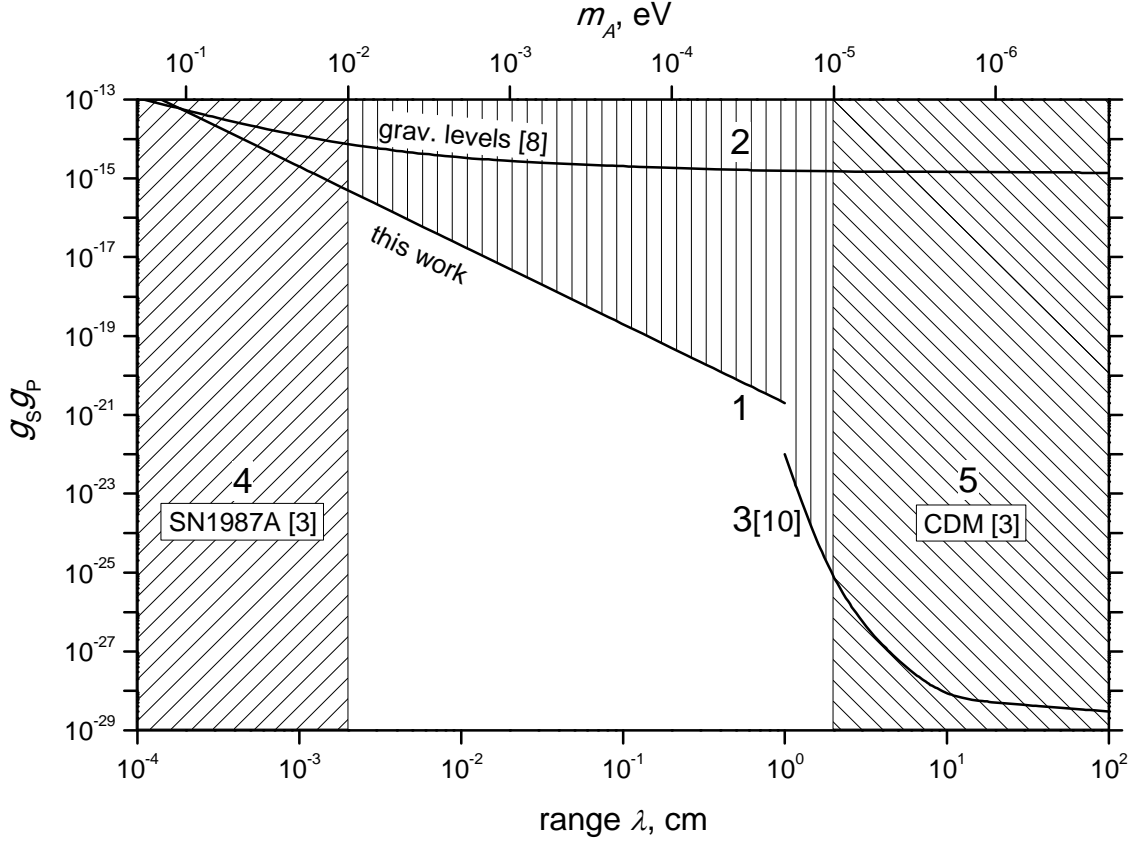

Figure 3: Constraints to the coupling constant product $g_s g_p$ of axion-like particles to nucleons as a function of the range $\lambda$ of the macroscopic force. On the upper horizontal axis the mass range of the axion-like particle is shown using the relation $\lambda = \hbar / m_A c$. Line 1: shift of resonance (this work); line 2: gravitational levels [8]; line 3 [10]; line 4: astrophysically excluded region of axion mass [3]; line 5: cosmologically excluded region in model of cold dark matter [3].

Using eq. (3) we can calculate the maximum strength of the pseudo-magnetic field on the surface of a thick copper wall still allowed by the limit on $g_s g_p$. For $\lambda$=1 μm one obtains about 0.1 mT while for $\lambda$=1 cm we have less than about 0.01 μT. These values illustrate the constraints on axion-like particles in electromagnetic units. A specific feature of the pseudo-magnetic field is that its source is non-polarized matter, in contra-distinction to ordinary magnetic fields. Observation of the pseudo-magnetic field requires the neutron as a "polarized detector".

It is also instructive to compare the obtained restrictions for CP-violating forces with constraints on deviations from the ordinary gravitational force in the CP-conserving sector. For this comparison we use the experimental constraints [9] obtained for the Yukawa-type parameterization of non-Newtonian forces,

$$V = -G_\text{N} \frac{m_1 m_2}{r}\left(1 + \alpha e^{-r/\lambda}\right) \quad (9)$$

where $G_\text{N}$ is Newton's constant, $m_1$ and $m_2$ are masses with center-of-mass separation $r$, and $\alpha$ is the relative strength of the non-Newtonian interaction with range $\lambda$. Constraints [9] for $\alpha$ are, for example, $\alpha < 10^{-3}$ for $\lambda = 1$ cm and $\alpha < 10^{10}$ for $\lambda = 3$ μm. If we parameterize the CP-violating forces expressed in eq. (2) and for $d \gg \lambda$ with respect to the gravitational force as in eq. (9), we find for the corresponding coupling strength $\alpha_\text{sp} < 10^2$ for $\lambda = 1$ cm and $\alpha_\text{sp} < 10^{14}$ for $\lambda = 1$ μm. Our constraints on CP-violating forces thus are about 5 orders of magnitude worse than constraints [9] on spin-independent non-Newtonian forces. However, the spin-dependent, CP-violating, and the spin-independent, CP-conserving forces represent independent sectors of macroscopic short-range interactions. One can see already from the form of the spin-dependent interaction that there is no net effect on an unpolarized probe, such that CP-violation may escape observation if one does not search for it with a spin probe.

We have presented first experiments which still offer large room for improvement. An improvement of the depolarization method might be possible with employment of materials with less magnetic impurities. The search for a resonance shift can be quite drastically improved. A sensitivity improvement down to $10^{-22}$ to $10^{-23}$ cm$^2$ for $g_s g_p \lambda^2$ might be within reach with the existing setup and UCN source. Of course all systematic effects have to be under control at the corresponding precision level. This will require another strategy which avoids hysteresis effects. The field reversal chosen in our first measurement will be replaced by a reversal of the mass-polarity of the trap at constant magnetic field [5]. Such efforts are under way.

The authors are grateful to the scientific committee and the administration of ILL for the possibility to carry out this experiment at the ultracold neutron source PF2 at the ILL reactor. The experiments were supported by the grant RFBR 07-02-00859.